# SAGE: Stealthy Attack GEneration for Cyber-Physical System

Michael Biehler, *Member, IEEE*, Zhen Zhong, and Jianjun Shi

*Abstract* — **Cyber-physical systems (CPS) have been increasingly attacked by hackers. Recent studies have shown that CPS are especially vulnerable to insider attacks, in which case the attacker has full knowledge of the systems configuration. To better prevent such types of attacks, we need to understand how insider attacks are executed. Typically, there are three critical aspects for a successful insider attack: (i) Maximize damage, (ii) Avoid detection, and (iii) Minimize the attack cost. This paper proposes a "Stealthy Attack GEneration" (SAGE) framework by formulizing a novel optimization problem considering these three objectives and the physical constraints of the CPS. By adding small worst-case perturbations to the system, the SAGE attack can generate significant damage, while remaining undetected by the systems monitoring algorithms. An efficient solution procedure for the nonconvex problem formulation is introduced. The proposed methodology is evaluated on several anomaly detection algorithms using different data modalities. The results show that SAGE attacks can cause severe damage while staying undetected and keeping the cost of an attack low.**

*Note to practitioners* — **This paper is motivated by the increasing cyber threats of insider attacks to CPS, which has unique characteristics and need further investigation. The proposed SAGE framework is able to generate worst-case attacks while staying undetected by the current systems monitoring algorithms and minimizing the attack cost. We model these objectives of an attacker as a novel optimization algorithm, which can be solved to achieve global optimality. Alternatively, off-the-shelf optimization heuristics such as genetic algorithms can also achieve efficient results for the nonconvex problem formulation. The SAGE formulation allows the flexible and holistic modeling of linear, hybrid, nonlinear and/or learning-enabled systems which have various applications in modern CPS. Ultimately, our method is intended to aid researcher and practitioners in the design and development of resilient CPS and detection algorithms.**

## I. Introduction

Cyber-PHYSICAL SYSTEMS (CPS) have been increasingly attacked by hackers [1]. However, the system integrity and security of CPS is of major importance, since any successful attack can lead to severe economic loss, equipment damage, or even loss of human life [2]. Understanding insider attacks is of great importance since they are becoming more frequent and more resources are needed to detect them [3, 4]. This indicates that the level of sophistication of attackers is increasing, and detection algorithms need to consider the complex structure of attacks executed by hackers who have full knowledge of the system.

Supervised machine learning techniques have achieved superior detection performance on existing types of attack generation schemes [5], [6], [7]. However, those type of supervised learning techniques require strong assumptions and can be considered as a best-case scenario for the defender of the system. In particular, historical training data needs to be available with labels of in-control (e.g., no attack) and attack conditions. Additionally, the current attack needs to come from the same generative process as the historical attacks.

In an unsupervised setting, control charts are commonly used to detect cyber-attacks in CPS [8]. However, in the research reported in this paper, we discover that if an attacker knows the current configuration of a CPS, most existing algorithms have vulnerabilities, which can be bypassed by the attackers. In a nutshell, the existing detection algorithms are based on very strong assumptions on the attack schemes, which may not mimic the behavior of an insider attacker. An effective detection algorithm requires the defenders to change perspective and "think like a hacker" to identify the weaknesses of a system and develop possible solutions to prevent intrusions.

Given the lack of holistic methods to generate insider attacks in CPS, we propose a general and comprehensive framework for "Stealthy Attack GEneration" (SAGE) in CPS. By formalizing a novel optimization problem, the SAGE framework considers the three main objectives of an attacker: (i) Maximize damage, (ii) Avoid detection, and (iii) Minimize the attack cost. By applying small, intentional, worst-case perturbations to the system variables, the SAGE attack will lead to unexpected and dangerous misbehavior of the system, while remaining undetected.

To show the generality of our approach, we generate stealthy attacks and validate the SAGE framework on two data modalities: image anomaly detection methods and functional curves from a hot rolling process simulated in MATLAB Simulink. In the image case studies, SSD [9] and CNN-LIME

M. Biehler is with the H. Milton Stewart School of Industrial and Systems Engineering (ISyE), Georgia Institute of Technology, Atlanta, GA 30332 USA (e-mail: michael.biehler@gatech.edu).

Z. Zhong is with the H. Milton Stewart School of Industrial and Systems Engineering (ISyE), Georgia Institute of Technology, Atlanta, GA 30332 USA (e-mail: zhongzhen@gatech.edu).

J. Shi is with the H. Milton Stewart School of Industrial and Systems Engineering (ISyE), Georgia Institute of Technology, Atlanta, GA 30332 USA (e-mail: jianjun.shi@isye.gatech.edu). Dr. Shi is the corresponding author.

are used as detection algorithms. In the case study of hot rolling process, seven commonly used detection algorithms are utilized, which include a Support Vector Machine (SVM), k Nearest Neighbor (kNN), Random Forest (RF), Bagging, Gradient Boosting Machine (GBM), Decision Tree (DT), and Deep Neural Network (DNN).

Our work contributes to the understanding of stealthy attacks in CPS. The results provide a case for the severe consequences of properly executed insider attacks. Furthermore, this research serves as a cornerstone for development of more effective detection algorithms and more resilient and robust CPS design. The remaining parts of this paper are organized as follows. In Section II, we review related literature to highlight the necessity of this research. In Section III, we present the mathematical description of a CPS, formulate the optimization problem, and propose the algorithm for solving this problem. In Section IV, the methods proposed in Section III are validated through case studies. Finally, Section IV concludes the paper.

## II. LITERATURE REVIEW

Due to the rise of smart manufacturing, CPS like power grids is increasingly exposed to cyber-attacks [10], [11], [12]. Attacks like the computer worm "Stuxnet" attacking Siemens industrial software in 2010 or the phishing attack on a German steel mill leading to severe equipment failures in 2014 are some of the most prominent examples for the vulnerability of CPS to cyber-attacks. Even though the field of information technology on cybersecurity is rapidly developing, the unique characteristics of CPS require specific attention [8]. CPS's have grown from stand-alone systems with little security protection to highly interconnected systems that can be easily targeted by attackers over the internet [13].

In general, the first step of an attacker is to gain knowledge of the system by identifying the network topology, software, critical targets, and monitoring schemes against cyber-attacks [14]. Then the first line of defense consisting of the firewall and an intrusion prevention system needs to be bypassed. After the attacker has full access to the CPS, the goal is to perturb the control systems and make as much damage as possible while staying undetected. The detection methods at the level of CPS like networks, systems and process data are considered the second line of defense.

### *A. Machine Learning Methods for CPS Attack Detection*

In recent years, multiple detection algorithms have developed for the second line of defense that utilize state-of-the-art machine learning classifiers [5], [15], [16], [17], [18], [19], [20]. Furthermore, machine learning methods are increasingly employed to CPS in many applications where they must perform complex tasks with a high degree of autonomy in uncertain environments. Traditional control design based on domain knowledge and analytical models are often impractical for tasks such as perception or control with ill-defined objectives. However, machine learning based techniques have demonstrated good performance for such difficult tasks, leading to the introduction of learning-enabled CPS. These methods mainly employ data-driven supervised machine learning methods like neural networks. However, little research in this domain has focused on modeling stealthy or covert attacks. Thus, the data generating process of any new attack needs to be the same, which is not reasonable in practice due to the vast number of attackers and methods in this field. Due to the drawbacks of classical machine learning techniques, we will introduce remedies that haven been proposed in the field of adversarial machine learning in the next subsection.

### *B. Adversarial Machine Learning*

Some prior work has addressed the problem of generating adversarial examples for machine learning systems [21], [22], [23], [24], but not in the context of fooling monitoring schemes in CPS by making very small perturbations of the control variables. In a cyber-attack CPS problem, additional the system model, the physical constraints, multiple monitoring algorithms and the attack cost need to be considered. Along this direction, [25] has develop an optimization framework for concealment attacks on reconstruction-based detectors.

The existing literature on generative models and adversarial neural networks in the field of cyber security [26], [27], [28], [29] mainly focuses on adversarial data, which could lead to wrong conclusions of a machine learning model. However, the literature of adversarial data has mainly focused on the image domain, and limited efforts have been made to generalize the concepts to a wide range of data modalities [30]. For example, Zizzo et al. (2020) [31] models an attacker to a system that monitors data via a Long Short-Term Memory (LSTM) time series model by optimizing $L_0$-norm perturbations to the system. Feng et al. (2017) [32] proposed a Generative Adversarial Network (GAN) based deep learning approach for stealthy attacks. However, this approach requires solving a nonconvex min-max optimization problem which in their setting is solved via Stochastic Gradient Descent (SGD). This approach suffers from several well-known theoretical and empirical problems in solving such saddle-point problems such as vanishing and exploding gradients. Additionally, this framework does not exploit any knowledge about the underlying system model or physical constraints. However, by utilizing the understanding of the generative model, more robust and reliable machine learning models can be developed.

### *C. State-Estimation Based Attacks and Defenses*

There is a large body of work in state-estimation techniques for cyber-physical intrusion detection in various safety-critical CPS such as industrial control systems or the power grid [33-36]. In particular, previous research has introduced a framework to generate integrity attacks by formalizing the adversary's strategy as a constrained control problem [37]. However, this method is only applicable to CPS that can be modeled as discrete linear time-invariant systems equipped with a Kalman filter, Linear Quadratic Gaussian (LQG) controller, and $\chi^2$ detector [37]. A constrained adversarial machine learning technique for CPS [38] has been proposed, but only considers the intrinsic constraints of the physical systems, which is only one of the objectives of an attacker. Furthermore, a wide variety of methods has been proposed to perturb the state estimation of a system. For example, a data framing attack on power system state estimation has been proposed by solving a quadratically constrained quadratic program [39]. For the smart grid state estimation, a malicious attack using a graph theoretic approach inducing $L_1$ norm pertubations has been proposed [40].

In the advent of learning-enabled CPS, the modeling of more holistic and stealthy attack is essential for the design of more resilient systems and detection algorithms. Therefore, the SAGE formulation seeks to provide a holistic generalization to the previously proposed attack schemes for a wide range of system models and monitoring schemes. This is facilitated by a flexible optimization framework that can be solved efficiently even for nonconvex settings.

## III. SAGE Methodology

This section first describes the system model used to model the dynamics of CPS. Afterwards, the SAGE framework is introduced, which considers the main objectives of an attacker consisting of maximizing the damage to the system while staying undetected and keeping the cost of an attack low. Finally, an efficient solution procedure for the nonconvex SAGE formulation is derived.

### *A. System Modeling*

The following section will describe the model used to characterize the system dynamics of CPS. For a general CPS, we will assume the following data scenario:

The process outputs $\boldsymbol{Y}_{k,t}$ from each subsystem *k* at time *t* can include multiple functional curves of the same length, images, structured point clouds, and categorical variables. Furthermore, we assume for each subsystem $i = 1,\dots,k$ there are $j = 1,..,q_i$ inputs at time *t* represented by $u_{ij,t}$. Contrary to an abundance of previous research in this field, we will assume that the effect of the inputs on the outputs can have a hybrid or nonlinear relationship. This allows more realistic modeling of complex CPS. This formulation can be adjusted appropriately according to the system model with the best fit to the historical data from a variety of potential models like linear regression, gaussian process model, or neural networks. Therefore, the system model can be represented as

$$\boldsymbol{Y}_{k,t} = \boldsymbol{B}_{k0} + \sum_{i=1}^{k}\sum_{j=1}^{q_i} g_{ij,t}\left(u_{ij,t}, \boldsymbol{\theta}\right) + \boldsymbol{E}_{k,} \tag{1}$$

where $g_{ij,t}(u_{ij,t}, \boldsymbol{\theta})(i = 1, \dots, k; j = 1, \dots, q_i)$ are some general function (e.g., linear, nonlinear, possibly varying with time) with the parameter vector $\boldsymbol{\theta}$, and $\boldsymbol{E}_k$ is a matrix containing the modeling error for each subsystem where every entry is a zero mean additive Gaussian noise with variance $\boldsymbol{\delta}^2{}_{E,k}$. The offset matrices are denoted by $\boldsymbol{B}_{k0}$.

As deep learning approaches are increasingly integrated into CPS (e.g., self-driving cars utilizing cameras for obstacle and traffic light detection), this general system model aims to unify a wide variety of models to model stealthy attacks in nonlinear settings. Furthermore, this general formulation also allows for the hybrid settings of linearized and nonlinear perception pipelines that are fused in a deterministic or stochastic manner.

### *B. Stealthy Attack GEneration (SAGE)*

Since 68% of organizations in a recent survey indicated that they are moderately to extremely vulnerable to insider attacks [3], we assume the worst-case scenario, where the attacker has full control and knowledge of the system. In particular, the attacker knows the process model and can inject data at any point and time.

Therefore, we will define a problem that fulfills the following three main objectives of an attacker:

i. *Maximize Damage:* The goal of an attacker is to cause damage to physical components such as machines, equipment, parts, assemblies, and products in CPS. Thus, the cyber attacker can cause devastating damage to CPS by increasing the wear, breakage, scrap, or any other changes to the original design [20].
ii. *Avoid Detection:* The aim of an attacker is to manipulate CPS in such a way that the altered control actions stay undetected. Most equipment has some hard-wired safety modes that will shut down the machines once they reach a safety relevant operating condition [41]. Therefore, staying undetected will directly contribute to the first objective to maximize damage. Once an intrusion is detected by the consecutive layers of cyber-attack detection, for example, on the firewall level, it would be hard to determine for the defender if and which parameters were actually changed by the intruder, which will cause even more downstream damage.
iii. *Minimize Attack Cost:* Attacking all control actions and replaying old data for all of them might be costly or complicated because different sensing data are saved in different databases and operating systems and many resources would be needed. Therefore, the attacker will want to keep the cost of an attack low.

Consequently, the attacker's optimization problem is formulated as Eq. 2, which exploits CPS system model and the weaknesses of the detection algorithm while considers the physical constraints of the system.

$$\min_{\boldsymbol{u}_t^A} -\left\| d_1(\boldsymbol{B}_{k0} + \sum_{i=1}^{k}\sum_{j=1}^{q_i} g_{ij,t}(u_{ij,t}^{IC}, \boldsymbol{\theta})) - d_2(\boldsymbol{B}_{k0} + \sum_{i=1}^{k}\sum_{j=1}^{q_i} g_{ij,t}(u_{ij,t}^{A}, \boldsymbol{\theta}))\right\|_p$$

$$s.t.$$

$$\|f(\boldsymbol{u}_t^{IC}) - f(\boldsymbol{u}_t^{A})\|_p \le \varepsilon_1$$
$$\|h(\boldsymbol{u}_t^{A}, \boldsymbol{u}_{t-1}^{A})\|_p \le \varepsilon_2$$
$$C(\boldsymbol{u}_t^{A}) \le \varepsilon_{3,} \tag{2}$$

where $d_1(\cdot)$ and $d_2(\cdot)$ denote a damage function corresponding to some undesirable outputs of a system given the in-control and attack control actions, respectively. Furthermore, $\boldsymbol{u}_t^A = \left(u_{11,t}, u_{12,t}, \dots, u_{21,t} \dots, u_{kq_i}\right)$ are the perturbed process inputs of all $k$ subsystems, and $q_i$ variables by the attacker, which should close to the in-control process inputs $\boldsymbol{u}_t^{IC} = \left(u_{11,t}^{IC}, u_{12,t}^{IC}, \dots, u_{21,t}^{IC} \dots, u_{kq_i,t}^{IC}\right)$. The distances are denoted in terms of the $\ell_p$-norm to allow for flexible modeling requirements. $\varepsilon_1$ denotes the maximal allowable distance between some general monitoring function $f(\cdot)$ applied to $\boldsymbol{u}_t^A$ and $\boldsymbol{u}_t^{IC}$; $\varepsilon_2$ denotes the maximal allowable distance between some general physical relationship $h(\cdot)$ of the attack at different time steps (e.g., $\boldsymbol{u}_t^A$ and $\boldsymbol{u}_{t-1}^A$**)** and $\varepsilon_3$ denotes the maximal allowable cost of an attack strategy $\boldsymbol{u}_t^A$. Using the Karush-Kuhn-Tucker (KKT) conditions, we can reformulate Eq. 2 to alleviate the burden of explicitly computing inequality constraints as follows:

$$\min_{\boldsymbol{u}_t^A} -\left\| d_1\left(\boldsymbol{B}_{k0} + \sum_{i=1}^{k}\sum_{j=1}^{q_i} g_{ij,t}(u_{ij,t}^{IC}, \boldsymbol{\theta})\right) - d_2\left(\boldsymbol{B}_{k0} + \sum_{i=1}^{k}\sum_{j=1}^{q_i} g_{ij,t}(u_{ij,t}^{A}, \boldsymbol{\theta})\right) \right\|_p + \lambda_1 \|f(\boldsymbol{u}_t^{IC}) - f(\boldsymbol{u}_t^{A})\|_p + \lambda_2 \|h(\boldsymbol{u}_t^{A}, \boldsymbol{u}_{t-1}^{A})\|_p + \lambda_3 C(\boldsymbol{u}_t^{A}), \tag{3}$$

where $\lambda_1$, $\lambda_2$ and $\lambda_3$ denote the Lagrange multipliers which corresponding to the constraints $\varepsilon_1$, $\varepsilon_2$ and $\varepsilon_3$, respectively.

The global minimum of the original constrained optimization problem (Eq. 2) corresponds to a saddle point in the Lagrangian function (Eq. 3), provided that the (necessary) regularity conditions of stationarity, primal feasibility, dual feasibility, and complementary slackness are satisfied. For a more detailed explanation of this widely used approach, interested readers are referred to [42]. We note that for nonconvex optimization problems, the Lagrange multipliers $\lambda_1$, $\lambda_2$ and $\lambda_3$, may not be unique. Therefore, we resort simultaneously solving for the optimal solution and the appropriate Lagrange multipliers by utilizing the Branch-and-Reduce framework introduced in Subsection C.

The detailed explanation of each term in Eq. 3 is as follows:

- The first term incorporates the first objective of the attacker, which is to maximize the damage to the system. This is equivalent to minimizing the negative of difference between the damage function $d(\cdot)$ for the in-control and attackers control actions respectively. If only the system output deviation is of concern, $d_1(\cdot)$ and $d_2(\cdot)$ reduces to the identify function. In cases where the state space has significant asymmetries, the $d(\cdot)$ functions can be defined as a (binary) mapping to a dangerous state.
- The second term ensures that the attack does not get detected. Depending on the detection algorithm, we choose a mapping function $f(\cdot)$ so attacker's control actions are close to "in-control" actions.
- The third term ensures that the physical constraints of the CPS are met via a function $h(\cdot)$, that maps the attacker's actions to the physical constraints. Control actions can only change within physical limits e.g., the magnitude of change in consecutive time steps should be small.
- The last term keeps the cost of attack low by considering how costly it is to attack a particular control action.

The system model as introduced in Subsection A is known in advanced or at least the predictions are accessible in a black box setting. The functions $f(\cdot)$ and $h(\cdot)$ are also known in advance. In Table I, several common monitoring statistics and physical constraints are introduced as guiding examples for the choice of $f(\cdot)$ and $h(\cdot)$.

TABLE I

EXAMPLES FOR AND THE CHOICE OF $f(\cdot)$ AND $h(\cdot)$

| **Monitoring Scheme** | $f(\cdot)$ | **Physical Constraint** | $h(\cdot)$ |
|---|---|---|---|
| X-bar & S Charts By default | Identity + Variance | Smooth changes over time | $u_{ij,t}^A - u_{ij,t-1}^A$ |
| Hotelling Control Chart | $T^2$ statistic | Sparse changes over time | $\left\|u_{ij,t}^A - u_{ij,t-1}^A\right\|_1$ |
| Kernel Methods (e.g., SVM or PCA) | Corresponding Kernel function | Limited variation patterns | $\left\|u_{ij,t}^A - u_{ij,t-1}^A\right\|_*$ |
| Gradient Boosting | Weighted sum of weak learners | Piecewise constant changes | Fused lasso penalty [43] |
| Neural Network Architectures | Inverse network function via back-propagation [44] | Variables within physically possible limits | $\left\|u_{ij,t}^A\right\|_2^2$ with appropriate Lagrange multiplier $\lambda_2$ |

$\|\cdot\|_*$ *denotes the nuclear norm.*

If the monitoring scheme or physical constraints are not known, the functions can be chosen as the identity and variance function by default as introduced in the steel rolling case study in Section 4.2.

Nonetheless, the main limitation of the SAGE attack is the assumption that $f(\cdot)$ and $h(\cdot)$ need to be known in advance in order to make a stealthy attack. If a defender constantly changes its monitoring algorithm or even system setup, it becomes very hard to exploit the vulnerabilities in a persistent manner. In particular, if the detector that monitors the system is not fully characterized, this attack framework might not lead to stealthy attacks. Another limitation is the possibly nonconvex formulation, which only under certain conditions has an optimality guarantee as discussed in the next section.

### *C. Solution Procedure*

The SAGE problem formulation is an inherently nonconvex and NP-hard problem. To make the SAGE framework applicable to wide range of general (nonconvex) functions the Branch-And-Reduce Optimization Navigator (BARON) algorithm is utilized to solve the nonconvex formulation to global optimum [45]. For simplicity of notation, we can express Eq. 2 as a general mathematical programming model.

$$\min f(\boldsymbol{x})$$

$s.t.$

$$g_1(\boldsymbol{x}) \le \varepsilon_1$$
$$g_2(\boldsymbol{x}) \le \varepsilon_2$$
$$g_3(\boldsymbol{x}) \le \varepsilon_3$$
$$\boldsymbol{x} \in \boldsymbol{X} \subseteq \mathbb{R}^n, \tag{4}$$

where $\boldsymbol{x} = \boldsymbol{u}_t^A \in \mathbb{R}^n$,

$$f(\boldsymbol{x}) = -\left\| d_1\left(\boldsymbol{B}_{k0} + \sum_{i=1}^{k}\sum_{j=1}^{q_i} g_{ij,t}(u_{ij,t}^{IC}, \boldsymbol{\theta})\right) - d_2\left(\boldsymbol{B}_{k0} + \sum_{i=1}^{k}\sum_{j=1}^{q_i} g_{ij,t}(u_{ij,t}^{A}, \boldsymbol{\theta})\right) \right\|_p : \mathbb{R}^n \to \mathbb{R},$$

$$g_1(\boldsymbol{x}) = ||f(\boldsymbol{u}_t^{IC}) - f(\boldsymbol{u}_t^A)||_p: \mathbb{R}^n \to \mathbb{R}^{m_1},$$
$$g_2(\boldsymbol{x}) = ||h(\boldsymbol{u}_t^A, \boldsymbol{u}_{t-1}^A)||_p: \mathbb{R}^n \to \mathbb{R}^{m_2},$$
$$g_3(\boldsymbol{x}) = C(\boldsymbol{u}_t^A): \mathbb{R}^n \to \mathbb{R}^{m_3}.$$

The output size of the nonconvex constraint functions is denoted by $m_1$, $m_2$ and $m_3$, respectively and $\boldsymbol{X}$ denotes a set of easy constraints to reduce the search space. For example, $\boldsymbol{X}$ could denote the $6\sigma$ limits of the attacked system levels, because any attack outside of those limits can very easily be detected. The standard Lagrangian subproblem of Eq. 4 is given in Eq. 3. However, for the dual approach to yield any computational advantage, the so-called Lagrangian subproblem must be much easier to solve than the primal problem. Therefore, we can define the Lagrangian subproblem as:

$$\inf_{x \in X} l'(\boldsymbol{x}, (\lambda_0, \lambda_1, \lambda_2, \lambda_3) = \inf_{x \in X}\{-\lambda_0 f(\boldsymbol{x}) - \lambda_1 g_1(\boldsymbol{x}) - \lambda_2 g_2(\boldsymbol{x}) - \lambda_3 g_3(\boldsymbol{x})\}, \quad (5)$$

where $(\lambda_0, \lambda_1, \lambda_2, \lambda_3) \leq 0$. The additional dual variable $\lambda_0$ homogenizes the problem and allows us to reformulate the SAGE attack into a unified BARON range-reduction problem. The constraints $\varepsilon_1, \varepsilon_2, \varepsilon_3$ enter the Lagrangian subproblem as $\lambda_1\varepsilon_1, \lambda_2\varepsilon_2$, and $\lambda_3\varepsilon_3$, respectively. Therefore, they are constants that do not alter the optimal solution and only need be considered in the Lagrangian master problem (Eq. 3). Assume that $b_0$ is an upper bound on the optimal objective function value of Eq. 2 and consider the following range-reduction problem:

$$h^* = \inf_{x, u_0, u_1, u_2, u_3} \{h(u_o, u_0, u_1, u_2, u_3) | f(\boldsymbol{x}) \leq u_0 \leq b_0, g_1(\boldsymbol{x}) \leq u_1 \leq \varepsilon_1, g_2(\boldsymbol{x}) \leq u_2 \leq \varepsilon_2, g_3(\boldsymbol{x}) \leq u_3 \leq \varepsilon_3, \boldsymbol{x} \in \boldsymbol{X}\}, \quad (6)$$

where $h$ is assumed to be some semi continuous function. Then Eq. 6 can be restated as

$$h^* = \inf_{x, u_0, u_1, u_2, u_3} h(u_0, u_1, u_2, u_3)$$
$$s.t.$$
$$-\lambda_0(f(\boldsymbol{x}) - u_0) - \lambda_1(g_1(\boldsymbol{x}) - u_1) - \lambda_2(g_2(\boldsymbol{x}) - u_2) - \lambda_3(g_3(\boldsymbol{x}) - u_3) \leq 0$$
$$(\lambda_0, \lambda_1, \lambda_2, \lambda_3) \leq 0$$
$$(u_0, u_1, u_2, u_3) \leq (b_0, \varepsilon_1, \varepsilon_2, \varepsilon_3)$$
$$\boldsymbol{x} \in \boldsymbol{X} \quad (7)$$

However, the computational complexity of Eq. 7 is the same as Eq. 5. Therefore, we lower bound $h^*$ with the optimal value of the following problem.

$$h_L = \inf_{x, u_0, u_1, u_2, u_3} h(u_0, u_1, u_2, u_3)$$
$$s.t.$$
$$\lambda_0 u_0 + \lambda_1 u_1 + \lambda_2 u_2 + \lambda_3 u_3 + \inf_{x \in X}\{-\lambda_0 f(\boldsymbol{x}) - \lambda_1 g_1(\boldsymbol{x}) - \lambda_2 g_2(\boldsymbol{x}) - \lambda_3 g_3(\boldsymbol{x})\} \leq 0$$
$$(\lambda_0, \lambda_1, \lambda_2, \lambda_3) \leq 0$$
$$(u_0, u_1, u_2, u_3) \leq (b_0, \varepsilon_1, \varepsilon_2, \varepsilon_3) \quad (8)$$

This domain reduction problem can be leveraged for efficiently solving the SAGE attack by restricting $h(u_0, u_1, u_2, u_3)$ to $a_0u_0 + a_1u_1 + a_2u_2 + a_3u_3$, where $(a_0, a_1, a_2, a_3) \geq 0$ and $(a_0, a_1, a_2, a_3) \neq 0$. Using Fenchel-Rockafellar duality, the BARON algorithm derived in [46] can be applied to iteratively obtain lower and upper bounds on the range-reduction problem of the SAGE attack formulation.

Branch- and Reduce (BARON) algorithm to solve SAGE attack

(0) **Initialize**: Set $K = 0, u_0^0 = a_0, u_1^0 = \varepsilon_1, u_2^0 = \varepsilon_2, u_3^0 = \varepsilon_3$

(1) **Solve the relaxed dual of Eq. 6:**

$$h_U^K = \max_{u_0, u_1, u_2, u_3} (\lambda_0 + a_0)b_0 + (\lambda_1 + a_1)\varepsilon_1 + (\lambda_2 + a_2)\varepsilon_2 + (\lambda_3 + a_3)\varepsilon_3 - z$$
$$s.t. \quad z \geq \lambda_0 u_0^k + \lambda_1 u_1^k + \lambda_2 u_2^k + \lambda_2 u_2^k, k = 0, \dots, K-1$$
$$(\lambda_0, \lambda_1, \lambda_2, \lambda_3) \leq -(a_0, a_1, a_2, a_3)$$

Let the solution be $(\lambda_0^K, \lambda_1^K, \lambda_2^K, \lambda_3^K)$

(2) **Solve the Lagrangian subproblem:**

$$\inf_{x, u_0, u_1, u_2, u_3} l'(\boldsymbol{x}, (\lambda_0^K, \lambda_1^K, \lambda_2^K, \lambda_3^K) = -\max_{x, u_0, u_1, u_2, u_3} \lambda_0^K u_0 + \lambda_1^K u_1 + \lambda_2^K u_2 + \lambda_3^K u_3$$
$$s.t. \quad f(\boldsymbol{x}) \leq u_0$$
$$g_1(\boldsymbol{x}) \leq u_1$$
$$g_2(\boldsymbol{x}) \leq u_2$$
$$g_3(\boldsymbol{x}) \leq u_3$$
$$\boldsymbol{x} \in \boldsymbol{X}$$

Let the solution be $(\boldsymbol{x}^K, u_0^K, u_1^K, u_2^K, u_3^K)$.

(3) **Augment and solve the relaxed primal problem:**

$$h_L^K = \min_{u_0, u_1, u_2, u_3} a_0u_0 + a_1u_1 + a_2u_2 + a_3u_3$$
$$s.t. \quad \lambda_0^k u_0 + \lambda_1^k u_1 + \lambda_2^k u_2 + \lambda_3^k u_3 + \inf_{x \in X} l'(\boldsymbol{x}, (\lambda_0^k, \lambda_1^k, \lambda_2^k, \lambda_3^k) \leq 0, k = 1, \dots, K$$
$$(u_0, u_1, u_2, u_3) \leq (b_0, \varepsilon_1, \varepsilon_2, \varepsilon_3)$$

(4) **Termination check:**

If $h_U^K - h_L^K \leq tolerance$

Stop

Else

Set K=K+1 and go to Step 1

In step 2 of the algorithm, $(\lambda_0, \lambda_1, \lambda_2, \lambda_3) \leq 0$ implies that $u_0^K = f(\boldsymbol{x}^K), u_1^K = g_1(\boldsymbol{x}^K), u_2^K = g_2(\boldsymbol{x}^K)$, and $u_3^K = g_3(\boldsymbol{x}^K)$.

Furthermore, the relaxations of the BARON framework enjoy quadratic convergence properties and are an efficient procedure for obtaining global optima to nonlinear programs [47]. In particular, the theorem for optimality-based range reduction [46] applies to the derived BARON algorithm for solving the SAGE attack:

**Theorem 1**. Suppose the Lagrangian subproblem in Eq. 5 is solved for certain dual multipliers $(\lambda_0, \lambda_1, \lambda_2, \lambda_3) \leq 0$. Then, for each $i$ such that $(\lambda_0^i, \lambda_1^i, \lambda_2^i, \lambda_3^i) \neq 0$, the cuts $g_p^i(\boldsymbol{x}) \geq (b_0 - \inf_x l(\boldsymbol{x}, \lambda_0, \lambda_1, \lambda_2, \lambda_3)/\lambda_p^i, \ p = 0,1,2,3)$ do not chop off any optimal solution of Eq. 4.

This implies that the solution will eventually converge to a global optimum even for nonlinear programs due to the quadratic convergence of the BARON algorithm. For a detailed discussion, related proofs, and generalizations we refer interested readers to [46].

For readers interested in generating SAGE attacks with no in-depth optimization knowledge, we recommend solving this problem using efficient heuristic methods such as the genetic algorithm (GA). When using heuristic methods with common software platforms, a few best practices should be considered. The attacks should be initialized with historic, in-control data. This will lead to much faster convergence. Furthermore, choosing upper and lower bounds within physical limits of the data (e.g., image pixel values from 0 to 255, system variables within $6\sigma$ limits) will reduce the probability of detection and drastically reduce the solution space of the problem. One drawback of this approach is, that heuristics do not provide any optimality guarantees. This however might not be of the utmost important for a persistent attacker of the system. An attacker only needs to achieve one bad system response to make damage, while in other circumstances the global optimality is much more desirable for example to achieve controllability and diagnosability. However, the choice of Lagrange multipliers of the SAGE formulation is crucial to the efficacy of the attack. Therefore, binary search can be adapted to find the optimal set of parameters for any arbitrary choice of algorithm.

Binary Search for Lagrange multiplier tuning of SAGE formulation

(1) **Input**:
Parameters $Y_{k,t}^{ref}$, $B_{k0}$, $g_{ij,t}$, $\boldsymbol{u}_t^A$, $\boldsymbol{u}_t^{IC}$
Attack Effectivity (AE) Thresholds $\alpha_1$
Average Perturbation (AP) Threshold $\alpha_2$
Attack Cost (AC) Threshold $\alpha_3$
Maximum iterations $max_{iter}$
(2) **Initialize**: $\lambda_{l,min} = 0$, $\lambda_{l,max} = 1$, $\lambda_l = \lambda_{l,min}$, $l = 1,2,3$
Counter C=0
(3) **Repeat SAGE attack**
(4) If $AE < \alpha_1$
$\lambda_{1,max} = \lambda_1$, $\lambda_{2,max} = \lambda_2$
$\lambda_1 = (\lambda_{1,min} + \lambda_{1,max})/2$, $\lambda_2 = (\lambda_{2,min} + \lambda_{2,max})/2$
C=C+1
(5) If $AP > \alpha_2$
$\lambda_{1,min} = \lambda_1$, $\lambda_{2,min} = \lambda_2$
$\lambda_1 = (\lambda_{1,min} + \lambda_{1,max})/2$, $\lambda_2 = (\lambda_{2,min} + \lambda_{2,max})/2$
C=C+1
(6) If $AC > \alpha_3$
$\lambda_{3,min} = \lambda_3$
$\lambda_3 = (\lambda_{3,min} + \lambda_{3,max})/2$
C=C+1
(7) If C> $max_{iter}$
Break
(8) Else if
Break
(9) End
(10) **End**

The binary search considers those three main objectives of the attacker and tunes the hyperparameters $\lambda_l, l = 1,2,3$ until the Attack Effectivity (AE), Average Perturbation (AP) and the Attack Cost (AC) are within prescribed limits. The Attack Effectivity can either be computed by the first SAGE term or by an attack specific metric considering the attacked system model. Similarly, the Average Perturbation can be derived from the second and third SAGE terms or from the defender's monitoring algorithm. The Attack Cost is directly calculated from the fourth SAGE term.

## IV. CASE STUDIES

In this section, we use two case studies to validate the SAGE methodology proposed in Section III. We will demonstrate how to use the proposed framework for functional curves as well as for image data.

### *A. Case Studies with Functional Curve Data*

To show the vulnerability of common CPS to stealthy attacks, a MATLAB Simulink testbed [48] for one stage plate rolling is used to show the devasting effect of small but worst-case perturbations to functional curves in CPS. The testbed uses a Multiple Input Multiple Output (MIMO) LQG regulator to control the horizontal and vertical thickness of a steel beam in a hot steel rolling mill.

In this case study, the SAGE formulation reduces to the following optimization problem.

$$\min_{\boldsymbol{u}_t^A} -\left\|\boldsymbol{Y}_t^{ref} - \boldsymbol{B}_0 - \textstyle\sum_{j=1}^4 \beta_j u_{ij}^A\right\|_2^2 + \lambda_1\|\boldsymbol{u}_t^{IC} - \boldsymbol{u}_t^A\|_2^2 + \lambda_2\|\boldsymbol{u}_t^A - \boldsymbol{u}_{t-1}^A\| + \lambda_3 C(\boldsymbol{u}_t^A)\,, \quad (9)$$

where $\boldsymbol{Y}_t^{ref}$ denotes the engineering specification of quality response and is a constant in this case. In this setting, $d_1(\boldsymbol{B}_{k0} + \sum_{i=1}^{k}\sum_{j=1}^{q_i} g_{ij,t}(u_{ij,t}^{IC}, \boldsymbol{\theta})) = \boldsymbol{Y}_t^{ref}$ and $d_1$ reduces to the identity function (i.e., $d_1(\cdot) = id(\cdot)$). Since the system response in this case is measured in terms of x- and y-axis thickness variation, the goal would be to have no variation so $\boldsymbol{Y}_t^{ref} = 0$. In this case $\boldsymbol{u}_t^{IC}$ is chosen as historic data of the same length as the attack to mimic a replay attack. Furthermore, the cost function as chosen as $C(u_{ij,t}^A)=\begin{cases}0, \; for\; j = 1,3 \\ 2, for\; j = 2,4\end{cases}$. This represents the fact that the roll gap ($j = 1,3$) is easy to attack while the roller force ($j = 2,4$) requires more efforts.

The optimization problem was solved using the proposed BARON algorithm for the SAGE formulation in Section III C. For better visualization, only the first 100 time-steps of the attack are visualized in the following figures. The attack clearly avoids detection by the x bar chart and only introduces minor perturbations to the control variables as visualized in Fig. 1.

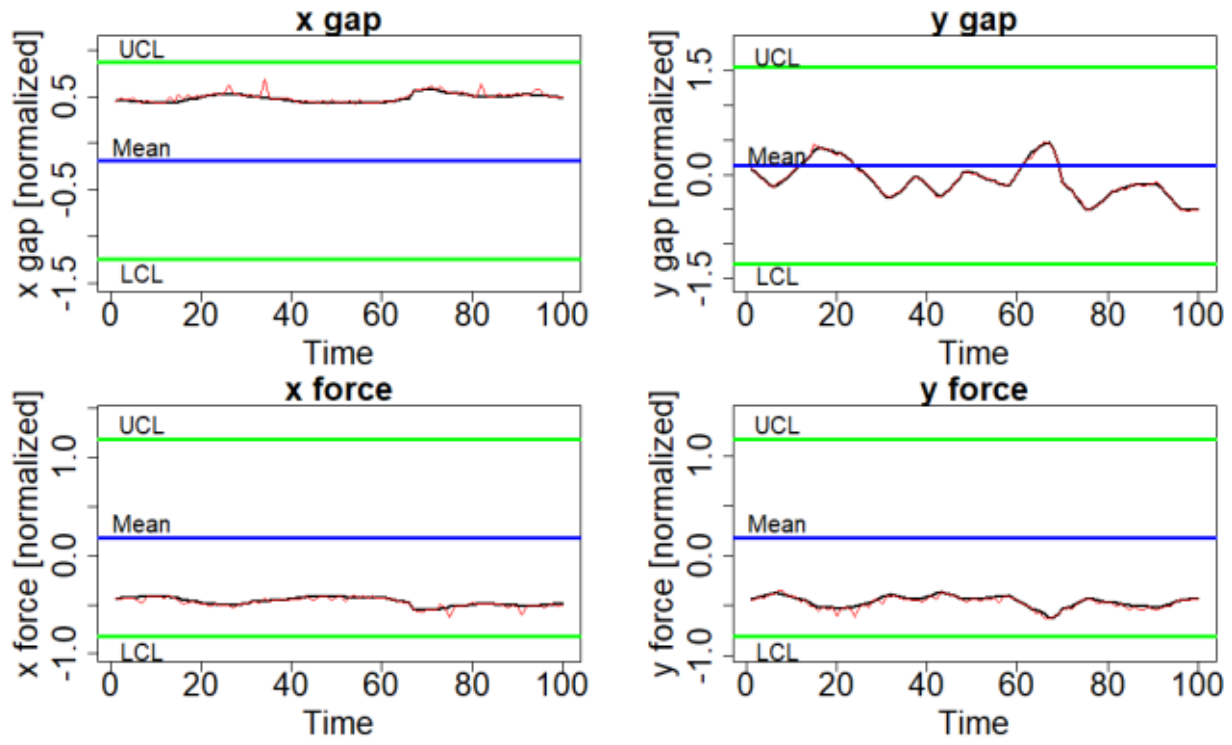

**Fig. 1.** Attackers' control actions (red) and in-control data (black) within 3$\sigma$ limits

On the other hand, the attack leads to maximal damage on the system response, which is far outside of the 3 sigma limits of the control chart (Fig. 2).

However, when looking at the difference of the correlation matrices in-control and attack data (Table II), we can see that the correlation structure is still different, which could lead to detection by the anomaly detection methods. The difference in correlation structure was calculated as the average of absolute difference of attack and in-control correlation matrix in a 100 time-steps sliding window over the entire dataset.

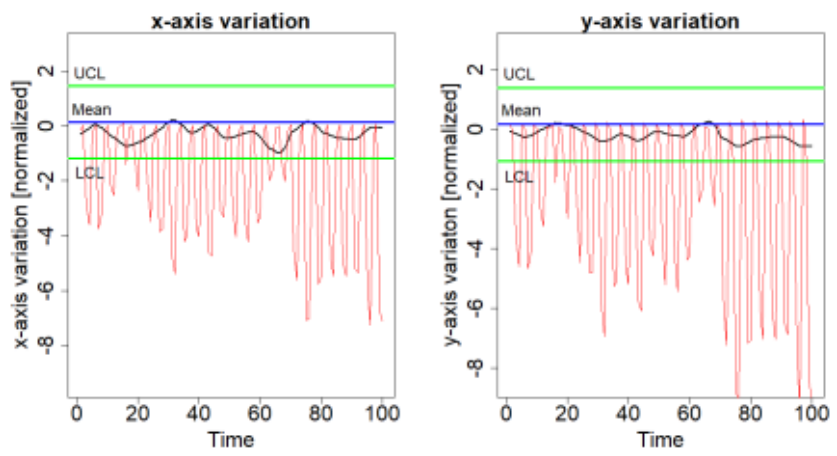

**Fig. 2.** System response (red) far outside of $3\sigma$ limits

TABLE II

ABSOLUTE DIFFERENCE BETWEEN IN-CONTROL AND ATTACK CORRELATION MATRIX, ATTACK 1 (EQ. 9)

| | x gap | x force | y gap | y force |
|---|---|---|---|---|
| x gap | 0 | | | |
| x force | 0.305 | 0 | | |
| y gap | 0.038 | 0.014 | 0 | |
| y force | 0.161 | 0.109 | 0.017 | 0 |

Therefore, the first penalty term of the SAGE formulation is extended to incorporate a similarity in terms of correlation structure as well.

$$\min_{\boldsymbol{u}_t^A} -\left\|\boldsymbol{Y}_t^{ref} - \boldsymbol{B}_0 - \sum_{j=1}^{4}\beta_j u_{ij}^A\right\|_2^2 + \lambda_{1,1}\|\boldsymbol{u}_t^{IC} - \boldsymbol{u}_t^A\|_2^2$$
$$+\lambda_{1,2}\left\|\sqrt{\frac{\sum_{m=t-n}^{t}\left(\boldsymbol{u}_m^{IC} - \overline{\boldsymbol{u}^{IC}}\right)^2}{n-1}} - \sqrt{\frac{\sum_{m=t-n}^{t}\left(\boldsymbol{u}_m^{A} - \overline{\boldsymbol{u}^{A}}\right)^2}{n-1}}\right\|_2^2$$
$$+\lambda_2\|\boldsymbol{u}_t^A - \boldsymbol{u}_{t-1}^A\|_2^2 + \lambda_3 C(\boldsymbol{u}_t^A), \quad (10)$$

where $\overline{\boldsymbol{u}^{IC}}$ and $\overline{\boldsymbol{u}^{A}}$ denote the process mean of the in-control data and attack data, respectively. Furthermore, a sliding window of $n = 100$ is used to compute the standard deviation in the added penalty term. To compute the initial sliding window, the first 100 time-steps of the new attack are initialized with previous results. To ensure the correlation structure was fully considered afterwards, the first 200 time-steps are discarded and not used by the attacker. As we can see from Fig. 3, that the signal of the in-control (black) and attack (red) are almost completely aligned. Therefore, they are indistinguishable for a human operator and hard to be detected by a control chart or other anomaly detection methods.

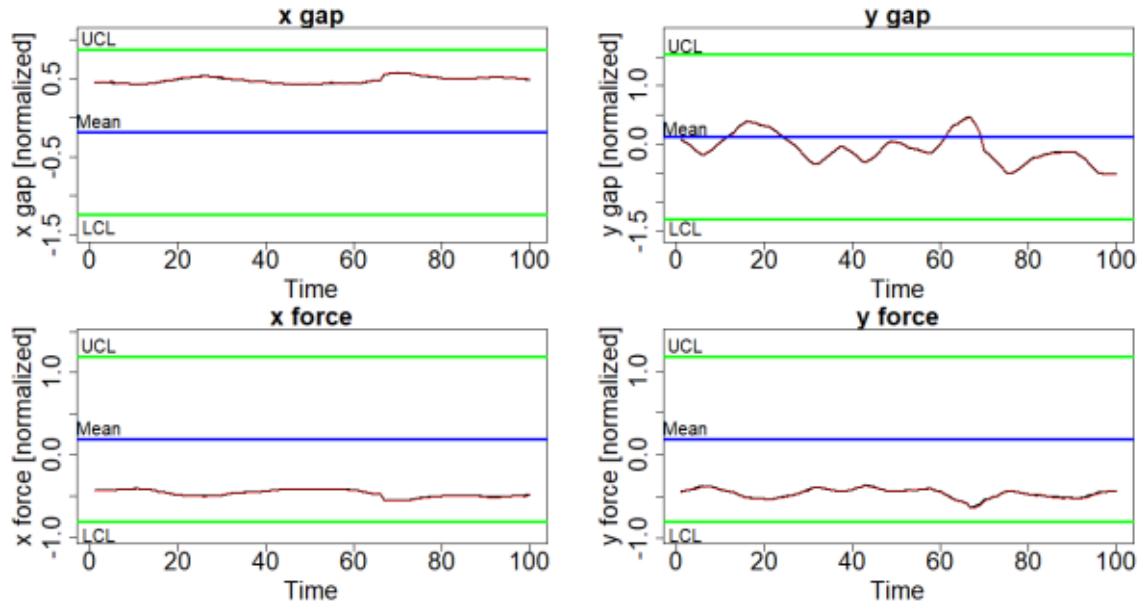

**Fig. 3.** Attackers' control actions (red) and in-control data (black) within $3\sigma$ limits

On the other hand, the SAGE attack can still cause a very abnormal system response (Fig. 4), which fulfills the main objectives of an attacker.

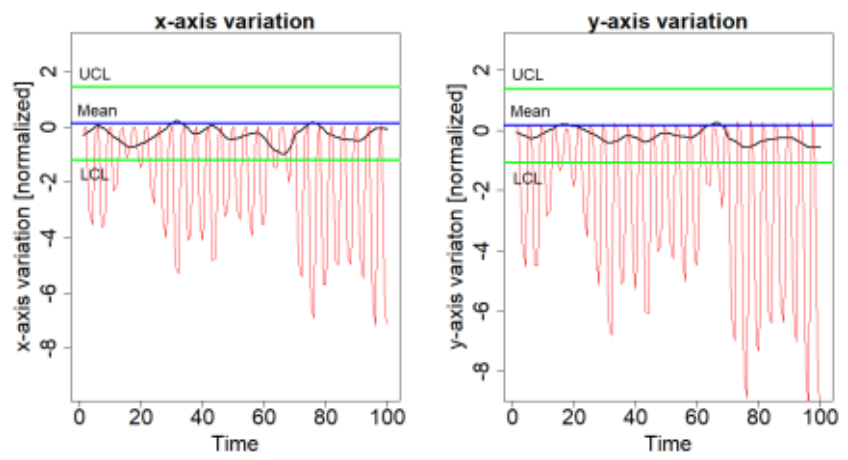

**Fig. 4.** System response (red) far outside of $3\sigma$ limits

After including a corresponding term into the SAGE formulation, the correlation structure of in-control and attack data shows no significant difference (Table III).

TABLE III

ABSOLUTE DIFFERENCE BETWEEN IN-CONTROL AND ATTACK CORRELATION MATRIX, ATTACK 2 (EQ. 10)

| | x gap | x force | y gap | y force |
|---|---|---|---|---|
| x gap | 0 | | | |
| x force | 0.007 | 0 | | |
| y gap | 0.014 | 0.016 | 0 | |
| y force | 0.009 | 0.002 | 0.023 | 0 |

To show that small perturbations of the control variables can lead to a large change of the system response, the Attack Effectivity (AE) and Average Perturbation (AP) are computed as follows.

- Attack effectivity $AE = \dfrac{\sum_{j=1}^{4}\left(\sum_{t=1}^{n}\left\|u_{j,t}^{IC} - u_{j,t}^{A}\right\|/n\right)}{\left(\sum_{j=1}^{4}\sigma_{u_j^{IC}}\right)}$
- Average Perturbation

$$AP = \frac{\sum_{t=1}^{n}\left\|\boldsymbol{Y}_t^{ref} - \boldsymbol{Y}_t^A\right\|/n}{\boldsymbol{\sigma}_Y},$$

where $n$ denotes the length of the attack, $\sigma_{u_j^{IC}}$ the in-control standard deviation of control variable $j$, $\boldsymbol{\sigma}_Y$ the in-control standard deviation of the system responses and $\boldsymbol{Y}_t^A$ is the resulting system response of the attack. Those metrics essentially measure the absolute distance of in-control and attack in terms the number of in-control standard deviations. The results are summarized in Table IV and show the significant decrease of perturbation from 0.129 (Attack 1, Eq. 8) to 0.037 (Attack 2, Eq. 9), while keeping the damage measured in terms of AE at a similar level.

TABLE IV

ATTACK EFFECTIVITY AND AVERAGE PERTURBATION OF SAGE ATTACKS

| | AE | AP |
|---|---|---|
| Attack 1 (Eq. 9) | 10.796 | 0.129 |
| Attack 2 (Eq. 10) | 10.636 | 0.037 |

To further evaluate the effectivity of the two proposed SAGE attacks, seven machine learning techniques commonly used in previous research for cyber-attack detection algorithms in CPS are evaluated for their effectiveness to detect stealthy attacks (Table V). The hyperparameters of the respective methods were tuned via grid search, to achieve the best possible detection results. In particular, a Support Vector Machine (SVM), k

Nearest Neighbor (kNN), Random Forest (RF), Bagging, Gradient Boosting Machine (GBM), Decision Tree (DT), and a Deep Neural Network (DNN) were used to classify the presence of an attack.

TABLE V

ACCURACY (ACC.), PRECISION (PREC.), RECALL (REC.) AND F1-SCORE OF DIFFERENT MACHINE LEARNING METHODS

| Method | Attack 1 (Eq. 9) | | | | Attack 2 (Eq. 10) | | | |
|---|---|---|---|---|---|---|---|---|
| | Acc. | Prec. | Rec. | F1 | Acc. | Prec. | Rec. | F1 |
| SVM | 0.671 | 0.797 | 0.674 | 0.635 | **0.623** | **0.784** | **0.625** | **0.563** |
| kNN | 0.458 | 0.445 | 0.460 | 0.419 | 0.394 | 0.394 | 0.394 | 0.394 |
| RF | 0.554 | 0.568 | 0.555 | 0.533 | 0.497 | 0.497 | 0.497 | 0.495 |
| Bagging | 0.545 | 0.547 | 0.546 | 0.542 | 0.420 | 0.420 | 0.420 | 0.420 |
| GBM | 0.574 | 0.575 | 0.575 | 0.573 | 0.497 | 0.748 | 0.501 | 0.334 |
| DT | 0.520 | 0.625 | 0.524 | 0.401 | 0.497 | 0.748 | 0.501 | 0.334 |
| DNN | **0.946** | **0.952** | **0.947** | **0.946** | 0.496 | 0.248 | 0.500 | 0.332 |

The results in Table V show that the DNN can detect the attacks generated by the first SAGE attack (Eq. 9) with very high accuracy (94.6%). In contrast to the other six methods, the DNN can capture a significant difference in the correlation structure as investigated in Table III. The high precision, recall, and F1-score of the DNN confirm these results. The other methods cannot capture the difference in the correlation structure and do not exhibit sufficient detection performance.

However, if the SAGE formulation (Attack 2) is adjusted to consider the correlation structure of the variables (Eq. 10), none of those six methods can achieve satisfactory detection performance. While the SVM performs the best, its detection accuracy of 62.3% is not sufficient for reliable and fast attack detection. Note, that in this setting the worst-case accuracy is 50% since the classifiers are trained on a balanced dataset (50% attack, 50% in-control). Therefore, in the worst case a random coin flip (i.e., attack, no attack) at each time point would result in a 50% accuracy. This example shows how flexible the SAGE formulation can be adjusted to make the existing detection algorithms not effective.

### *B. Case Studies with Image Data*

In this subsection, we will provide a generalization of the SAGE attack to learning-enabled CPS utilizing two state-of-the-art anomaly detection algorithms. Another goal of this case study is to illustrate the potential of the SAGE framework on other data formats, in particular image data. This provides a generalization to the previous research in the field of adversarial examples for machine learning. Furthermore, we provide a case for the severe consequences of small but intentional perturbations to control variables on image responses in CPS. Therefore, we will attack both the smooth spare decomposition (SSD) method [9], which is a benchmark image denoising and anomaly detection algorithm in the field of manufacturing, and a CNN-LIME, which is a state-of-the-art method in the field of classification and object detection.

The dataset used for both attacks is the Northeastern University (NEU) surface defect database [49], [50], [51], which contains six typical surface defects of hot-rolled steel strips. The dataset includes 1,800 grayscale images, with 300 samples of each of the six different surface defects (i.e., rolled-in scale (RS), patches (Pa), crazing (Cr), pitted surface (PS), inclusion (In) and scratches (Sc)).

#### *1. SAGE Attack on SSD Method*

Firstly, we attack the SSD method [9], which decomposes an image into three components: A smooth image background, the sparse anomalous regions, and the random noises, as illustrated in Fig. 5.

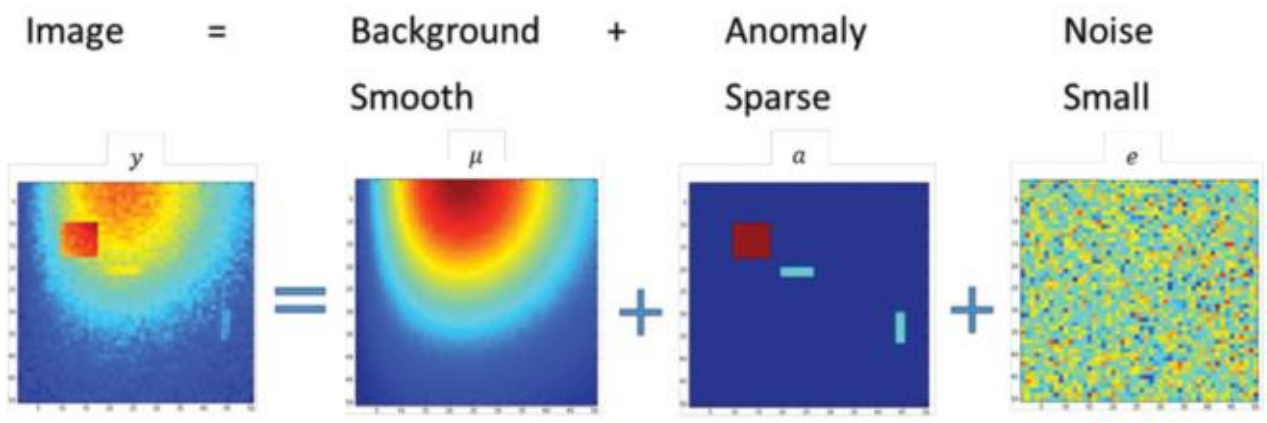

**Fig. 5.** Decomposition of image into background, anomaly, and noise [9]

The goal of the attack is to add small perturbations to the image, which are indistinguishable from the original image for the human eye and cannot be detected by designated detection algorithms. However, they will lead to a bad system response. In this case, the system response is the anomaly region, and we want to change the anomaly region as much as possible. When decomposing the image into background, anomaly, and noise via SSD, we want to detect the anomalies in different regions than where they actually are. This means, when the operators try to fix the problem, they will draw a wrong conclusion regarding the root cause of the anomalies and make the damage even worse by taking the wrong actions. In this circumstance, the SAGE attack formulation reduces to the following optimization problem.

$$\min_{\boldsymbol{y}_t^A} - \|\boldsymbol{\theta}_\alpha - \boldsymbol{\theta}_\alpha^{SSD}(\boldsymbol{y}_t^A)\|_F^2 + \lambda_1 \left\|\boldsymbol{y}_t^{original} - \boldsymbol{y}_t^A\right\|_F^2 + \lambda_3 C(\boldsymbol{y}_t^A) + \lambda_3 C(\boldsymbol{y}_t^A), \quad (11)$$

where $\boldsymbol{y}_t^A$ denotes the image that the attacker will inject into the system at time $t$, $\boldsymbol{\theta}_\alpha$ denotes the fixed and known anomaly region, $\boldsymbol{\theta}_\alpha^{SSD}$ is a function of $\boldsymbol{y}_t^A$ and denotes the extracted anomaly region from the attacker's image. The goal of the attacker is to maximize the damage by letting $\boldsymbol{\theta}_\alpha^{SSD}$ be as far away as possible from the ground truth anomaly $\boldsymbol{\theta}_\alpha$. Furthermore, to not get detected the attackers' image should be close to the original image before the attack $\boldsymbol{y}_t^{original}$. Since the monitoring of a process usually consists of streaming data from each time step $t$, the added perturbations in consecutive time steps should not be too different since this might be physically impossible. Furthermore, extreme changes over time might alert appropriate detection algorithms and lead to detection. This behavior is enforced by the second term in the formulation. The difference of the first three terms is calculated in terms of the squared Frobenius norm. Additionally, since the cost increases with the number of pixels attacked in an image, the cost function is chosen as the $l_1$-norm to induce sparsity and attack as few pixels as possible.

The SAGE attack on SSD was solved using the BARON framework introduced in Section III C.

As shown in Fig. 6, the image before and after the attack are almost indistinguishable to the human eye.

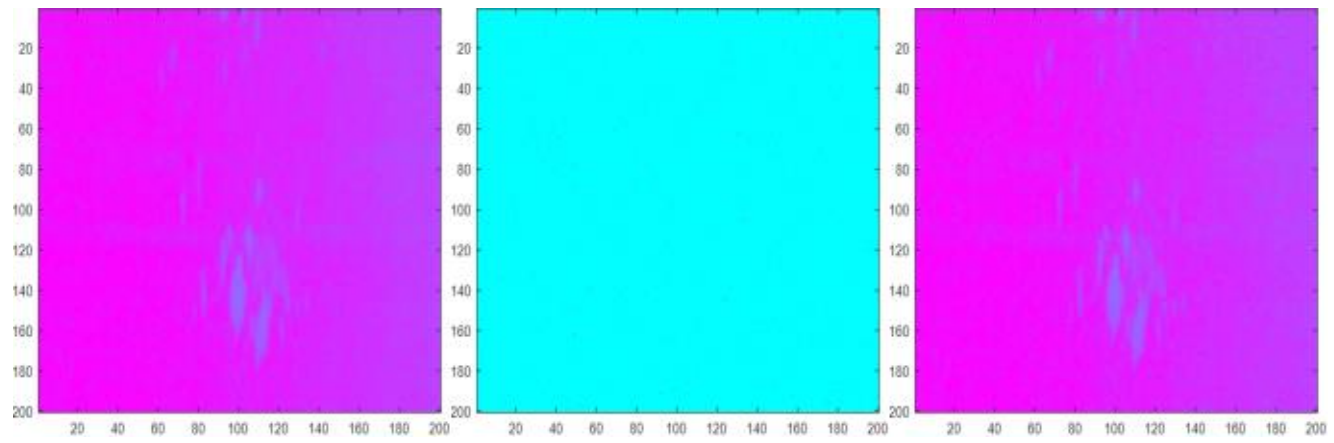

**Fig. 6.** Original image (left), added perturbations (middle) and attacked image (right) of examplary steel surface defect

On the other hand, the outputs of the SSD algorithm before and after the attack are quite different (Fig. 7). After the attack, the false alarm rate has increased significantly since many regions are now identified incorrectly as surface defects.

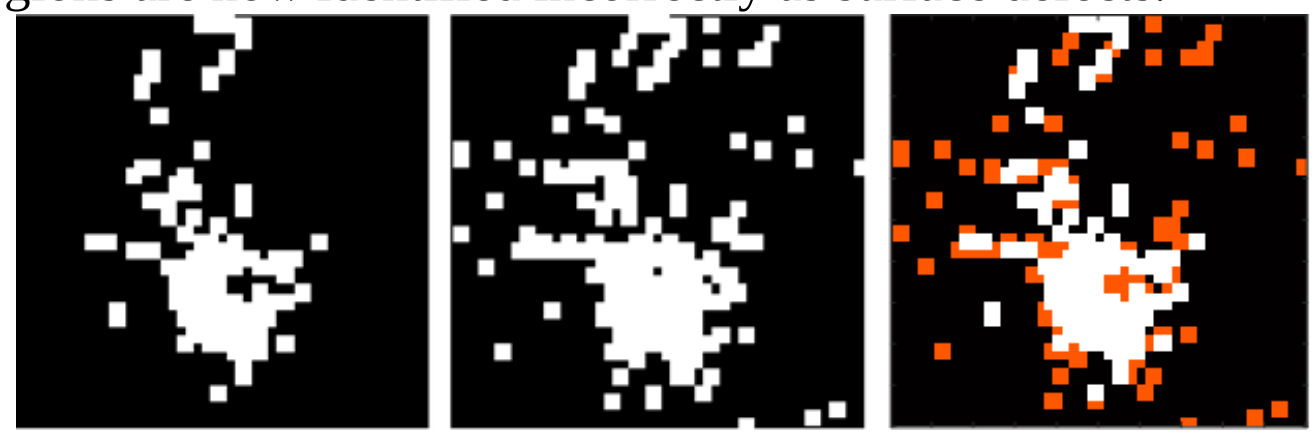

**Fig. 7.** Examplary recovered anomaly using SSD from the original image (left) anomaly from the attacked image (middle) and difference between anomaly region of original and attack image in red (left)

To show the generality of the SAGE formulation in attacking multiple classes of anomalies, the entire data set of 1,800 images is selected and the following metrics are defined corresponding to the objectives of the attacker.

- Attack effectivity $AE = \frac{\sum \mathbf{1}_{>0}\left(\left|\boldsymbol{\theta}_\alpha^{original} - \boldsymbol{\theta}_\alpha^{A}\right|\right)}{\mathbf{1}_{>0}(\boldsymbol{\theta}_\alpha^{A})}$
- Average Pixel Perturbation

$$APP = \frac{\sum_{k=1}^{n}\sum_{l=1}^{m}\left|Y_{kl}^{original} - Y_{kl}^{A}\right|}{n \cdot m \cdot 255},$$

where n and m denote the height and width of the image, respectively. In this case study the images have the size $n = m = 200$. The larger the attack effectivity, the more damage the attacker can do to the anomaly region; and the smaller the average pixel perturbation, the closer the attacked image will be to the original image. Note that the APP is scaled by 255 to account for the range of the pixel intensity values from 0 to 255. The averaged results of those metrics for the 1,800 images are shown in Table VI.

TABLE VI

ATTACK EFFECTIVITY AND AVERAGE PIXEL PERTURBATION OF SAGE ATTACK APPLIED TO SSD

| | AEE | APP |
|---|---|---|
| SAGE Attack | 40.534% | 0.0482 |

As we can see from the results of the surface defects, after applying small but intentional perturbations via the SAGE framework, the SSD algorithm can be fooled by falsely adding and/or deleting anomaly regions, while generating an attack image that is virtually indistinguishable for the human eye. This case study shows the generality of the SAGE framework when applied to image data even for complex anomaly detection algorithms like SSD, which utilizes advanced optimization techniques. Therefore, our proposed framework can easily be adapted for other image anomaly detection methods as long as the detection algorithms parameters are explicitly known or at least predictions from the detection algorithm can be accessed in a black box manner.

*2. SAGE Attack on CNN-LIME*

Furthermore, a Convolutional Neural Network in combination with Local Interpretable Model-Agnostic Explanations (CNN-LIME) [52] is attacked. LIME explains the prediction of any classifier by treating it as a black box model and learning an interpretable model locally around the prediction. LIME finds the region of an image that led to the classification of that image to a particular class. In view of this fact, it is related to object detection algorithms that locate objects of interest in an image by predicting a boundary around the object. Based on previous research [53], [23], [54], [55], object detection algorithms are much more difficult to attack [55]. Therefore, attacking CNN-LIME will demonstrate the immense capabilities of the proposed SAGE formulation in attacking a wide range of algorithms.

To obtain a good classification model, transfer learning with weights from the MobileNet is utilized. An 99.9% model accuracy can be achieved by initializing the CNN architecture with those weights and fine tuning it on the NEU surface detection dataset. The corresponding confusion matrix is visualized in Fig. 8.

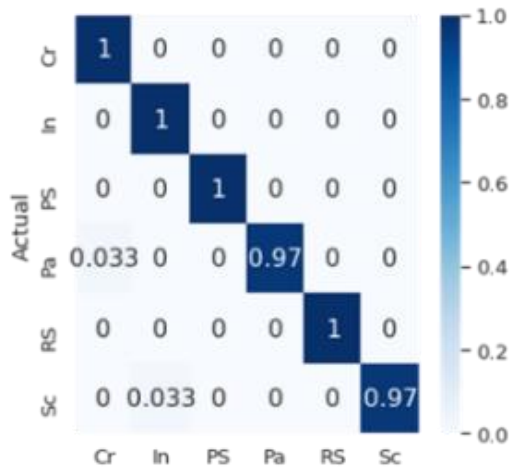

**Fig. 8.** Confusion Matrix of CNN Model

Afterwards the LIME algorithm is utilized to explain the predictions of the CNN model and identify the anomaly regions in the images.

The SAGE formulation is adopted as follows in this setting.

$$\begin{aligned} \min_{\boldsymbol{y}_t^A} & -\left\|\boldsymbol{\theta}_t^{original} - \boldsymbol{\theta}^{CNN}(\boldsymbol{y}_t^A)\right\|_F^2 \\ & -\lambda_0\left\|\boldsymbol{\xi}^{original} - \boldsymbol{\xi}^{LIME}(\boldsymbol{y}_t^A)\right\|_F^2 \\ & +\lambda_{1,1}\left\|\boldsymbol{y}_t^{original} - \boldsymbol{y}_t^A\right\|_F^2 \\ & +\lambda_{1,2}\left\|\mathrm{vec}\left(\boldsymbol{y}_i^{original} - \boldsymbol{y}_i^A\right) \cdot \mathcal{D} \cdot \mathrm{vec}\left(\boldsymbol{y}_i^{original} - \boldsymbol{y}_i^A\right)\right\|_F^2 \\ & +\lambda_3 C(\boldsymbol{y}_i^A) \end{aligned} \quad (12)$$

where $\boldsymbol{\theta}$ denotes the predicted class probabilities and $\boldsymbol{\xi}$ is the explanation produced by LIME for the class predictions. The intuition of this attack is to misclassify the anomaly images, while keeping the attacker's image close to the original image and changing the explanatory region away from the original one to make the attackers malicious class prediction seem legitimate. Furthermore, the changes in the image should be smooth to preserve the spatial dependencies to avoid detection. Therefore, the smoothness penalty $\lambda_{1,2}\left\|\mathrm{vec}\left(\boldsymbol{y}_t^{original} - \boldsymbol{y}_t^A\right) \cdot \boldsymbol{\mathcal{D}} \cdot \mathrm{vec}(\boldsymbol{y}_t^{original} - \boldsymbol{y}_t^A)\right\|_F^2$ is applied, where

$$\boldsymbol{\mathcal{D}} = \begin{bmatrix} 1 & -1 & & & \\ -1 & 2 & -1 & & \\ & \ddots & \ddots & \ddots & \\ & & -1 & 2 & -1 \\ & & & -1 & 1 \end{bmatrix}$$

is the second-order smoother that applies to the vectorized difference between the original and the attacker's image.

Similar to the image attack on the SSD algorithm, the attacker's image can hardly be distinguished from the original one as shown in Fig. 9.

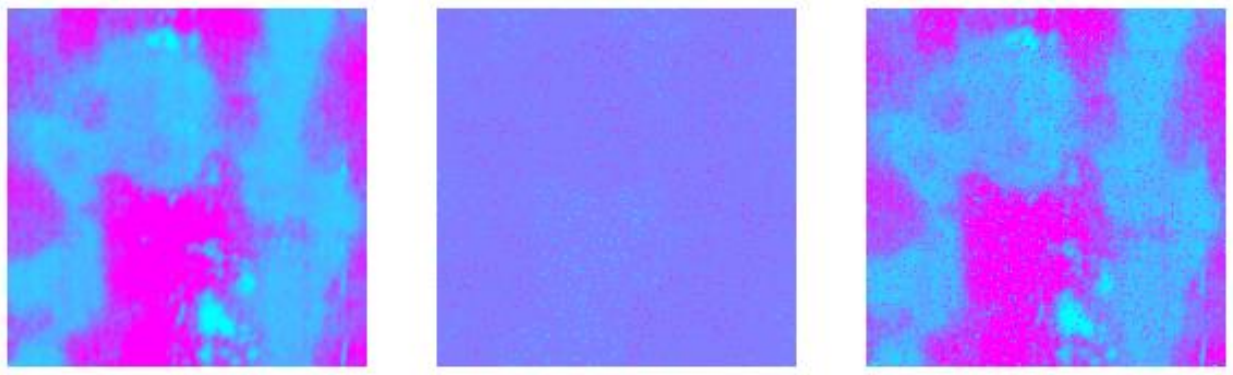

**Fig. 9.** Original image (left), added perturbations (middle) and attacked image (right) of examplary surface defect

On the other hand, an exemplary classification result changes significantly as shown in Table VII.

TABLE VII

EXEMPLARY CLASSIFICATION RESULTS BEFORE AND AFTER ATTACK

| Class Label | RS | PS | Cr | Pa | In | Sc |
|---|---|---|---|---|---|---|
| Before Attack | 0.001 | 0.000 | 0.002 | **0.996** | 0.001 | 0.001 |
| After Attack | 0.001 | 0.012 | 0.179 | 0.195 | **0.514** | 0.099 |

*The highest class probability in **bold**.*

In this attack formulation, the goal was to misclassify a given true process anomaly class as any of the remaining five class labels. From the example in Table VII, we can see that the correct class patches (Pa) is identified with very high confidence (99.6%) before the attack. After the attack, the probability of the correct class reduces to 19.5% and the class inclusion (In) was chosen with highest confidence (51.4%). Any other process anomaly class can be attacked in a similar fashion as summarized for the 1,800 images in the dataset in Table VIII.

If the attacker not only wants to misclassify the anomalies but also assigns the picture to a specific prescribed class, the first penalty term in Eq. 12 can be adjusted accordingly. The goal of the attack was also to change the explanatory region derived via LIME as far as possible from the original one to avoid any suspicion and justify the differently classified anomaly after the SAGE attack on the image. Fig. 10 shows an example of the severe change in explanatory region after the attack.

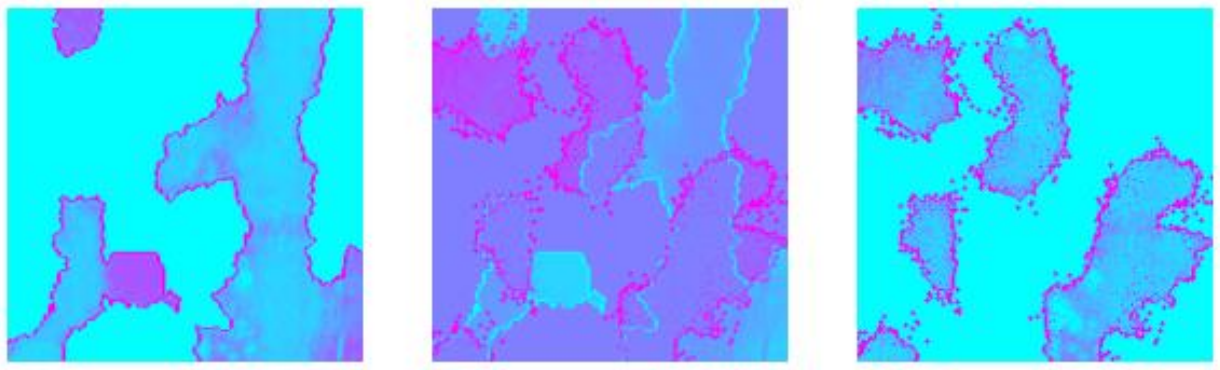

**Fig. 10.** Original explanatory region computed via LIME (left), change in explanatory region (middle) and attacked explanatory region (right) of exemplary surface defect

The small pixels around the identified regions after the attack coincide with the inclusion anomaly, which has the highest-class probability after the attack. This will avoid detection by the defender while leading to wrong conclusions about the underlying process anomaly.

The SAGE attack was applied to the entire dataset of 1,800 images. The attack metrics for those attacks are as follows:

- The change of classification is denoted as the Ratio of Attacked to Clean correct class Accuracy (RACA) as follows: $RACA = \frac{1}{n}\frac{\sum_{i\in Attack} L(y_i^A|\boldsymbol{Y}_i)}{\sum_{i\in Original} L(y_i^{original}|\boldsymbol{Y}_i)}$, where L denotes the accuracy loss of a single picture $\boldsymbol{y}_i$ with true class $\boldsymbol{Y}_i$ and n is the number of image samples. Note, a smaller score indicates a better attack.
- The change in the LIME explanatory region is denoted by the attack effectivity (AE) as defined earlier.
- The attacker's perturbation to the input image is denoted by the average pixel perturbation (APP) as defined earlier.

The averaged results for the entire dataset are reported in Table VIII.

TABLE VIII

AVERAGE ATTACK METRICS OF SAGE ATTACK APPLIED TO CCN-LIME

| | RACA for CNN | AE for LIME | APP |
|---|---|---|---|
| SAGE Attack | 29.526% | 69.534% | 0.0716 |

The results show the large effectiveness of the general SAGE attack on a large number of image classification results computed via CNN-LIME.

In summary, the SSD algorithm is much more vulnerable to perturbations than CNN-LIME. The SSD attacks exploits very few weak spots in the image and changes the pixel value significantly to destroy the smoothness of the background. The CNN-LIME attack has a slightly higher APP of 0.0716. To change the classification result to a different class of only six possible ones, a much large number of pixels needs to be attacked. Therefore, the CNN attack is a much more challenging task. However, the SAGE formulation can exploit the weaknesses of both SSD and CNN-LIME very effectively, while not utilizing any knowledge about the specific parameters and weaknesses of the two respective algorithms. In view of this fact, the SAGE attack provides an effective generalization for existing adversarial example generation schemes in the setting of a black-box attacks.

Even in the case of black-box attacks, where the detection algorithm is not known to the attacker, the proposed SAGE framework can cause severe damage to a system while staying undetected by commonly used machine learning classifiers. This provides a strong case for the generality and effectivity of the proposed framework, which can not only exploit weaknesses of particular algorithms through its flexible formulation, but also make replay non-essential for effective attacks by mimicking the normal operating conditions.

## V. Conclusion

We have introduced a holistic framework for attack generation in CPS (SAGE), which incorporates the three main objectives of an attacker (maximize damage, avoid detection, minimize the attack cost) and the physical constraints of the CPS. By solving the proposed optimization problem, SAGE can generate stealthy worst-case perturbations that fulfill all the objectives of the insider attacker. An efficient algorithm with convergence guarantees has been developed for solving this nonconvex optimization problem. This framework provides a general formulation to target any algorithm used for cyber-attack detection in CPS as long as the formulation is known to the attacker. Even if no specific knowledge of the detection method is available (black-box attack), state-of-the-art machine learning techniques can be fooled by mimicking the normal operating conditions as verified in the case studies.

This research contributes to the understanding of stealthy attacks in CPS. As our research shows, properly executed insider attacks, that mimic the same generative process as normal, in-control data, cannot be handled by many existing anomaly detection methods. This implies that future detection algorithms should not rely on readily available anomaly detection techniques but study the stealthy and adversarial behavior of cyber-attacks.

The SAGE framework is intended as a new benchmark for researchers and practitioners for the design of more efficient detection algorithms and more robust and resilient CPS design.

## Acknowledgment

This research is funded by the National Science Foundation Award ID 2019378.

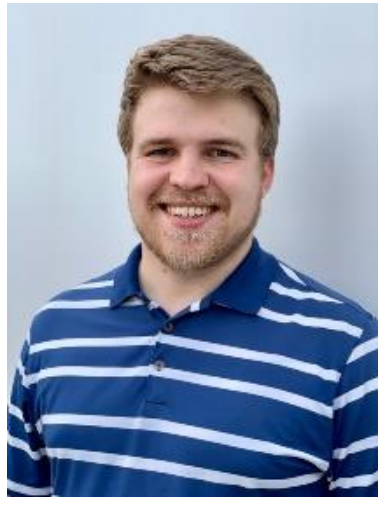

**Michael Biehler** (Member, IEEE) received his B.S. and M.S in Industrial Engineering and Management from Karlsruhe Institute of Technology (KIT) in 2017 and 2020, respectively. Currently, he is a Ph.D. student in the Stewart School of Industrial and Systems Engineering, Georgia Institute of Technology. His research rests at the interface between machine learning and cyber physical (manufacturing) systems, where he aims to develop methods for cyber security, monitoring, prognostics, and control. He is a member of ASME, ENBIS, IEEE, IISE, INFORMS and SIAM.

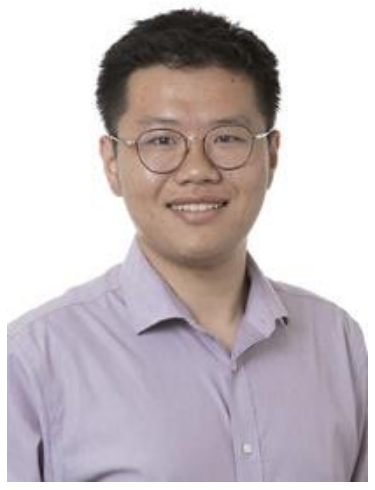

**Zhen Zhong** received the B.S. degree in Electrical Engineering from University of Science and Technology of China, Anhui, China, in 2017. Currently, he is a Ph.D. student at the H. Milton Stewart School of Industrial and Systems Engineering, Georgia Institute of Technology. His research interests include the process control and high dimensional data analytics in semiconductor manufacturing.

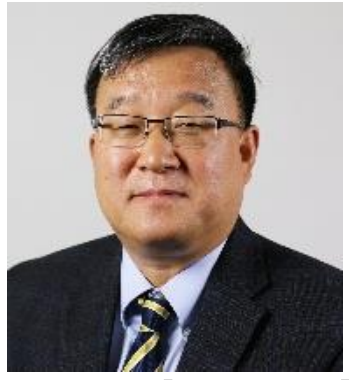

**Dr. Jianjun Shi** received the B.S. and M.S. degrees in automation from the Beijing Institute of Technology in 1984 and 1987, respectively, and the Ph.D. degree in mechanical engineering from the University of Michigan in 1992.
Currently, Dr. Shi is the Carolyn J. Stewart Chair and Professor at the Stewart School of Industrial and Systems Engineering, Georgia Institute of Technology. His research interests include the fusion of advanced statistical and domain knowledge to develop methodologies for modeling, monitoring, diagnosis, and control for complex manufacturing systems.
Dr. Shi is a Fellow of the Institute of Industrial and Systems Engineers (IISE), a Fellow of American Society of Mechanical Engineers (ASME), a Fellow of the Institute for Operations Research and the Management Sciences (INFORMS), an elected member of the International Statistics Institute (ISI), a life member of ASA, an Academician of the International Academy for Quality (IAQ), and a member of National Academy of Engineers (NAE).